# Bias reduction method for prior event rate ratio, with application to emergency department visit rates in patients with advanced cancer

Xiangmei Ma [1], Chetna Malhotra [2,3], Eric Andrew Finkelstein [2,3], Yin Bun Cheung [1,3,4]

1. Centre for Quantitative Medicine, Duke-NUS Medical School, National University of Singapore, 8 College Road, Singapore 169857
2. Lien Centre for Palliative Care, Duke-NUS Medical School, National University of Singapore, 8 College Road, Singapore 169857
3. Programme in Health Services & Systems Research, Duke-NUS Medical School, National University of Singapore, 8 College Road, Singapore 169857
4. Center for Child, Adolescent and Maternal Health Research, Tampere University, Tampere, FIN-33014, Finland

\* Corresponding author:

Dr Xiangmei Ma, Centre for Quantitative Medicine, Duke-NUS Medical School, 8 College Road, Singapore 169857

Email: xiangmei.ma@duke-nus.edu.sg

**Abstract**

**Objectives**:

Prior event rate ratio (PERR) is a promising approach to control confounding in observational and real-world evidence research. One of its assumptions is that occurrence of outcome events does not influence later event rate, or in other words, absence of "event dependence". This study proposes, evaluates and illustrates a bias reduction method when this assumption is violated.

**Study Design and Setting**:

We propose the conditional frailty method for implementation of PERR in the presence of event dependence and evaluate its performance by simulation. We demonstrate the use of the method with a study of emergency department visit rate and palliative care in patients with advanced cancer in Singapore.

**Results**:

Simulations showed that, in the presence of negative (positive) event dependence, the crude PERR estimate of treatment effect was biased towards (away from) the null value. The proposed method successfully reduced the bias, with median of absolute level of relative bias at about 5%. Dynamic random-intercept modelling revealed positive event dependence in emergency department visits among patients with advanced cancer. While conventional time-to-event regression analysis with covariate adjustment estimated higher rate of emergency department visits among palliative care recipients (HR=3.61, P<0.001), crude PERR estimate and the proposed PERR estimate were 1.45 (P=0.22) and 1.22 (P=0.57), respectively.

**Conclusions**:

The proposed bias reduction method mitigates the impact of violation of the PERR assumption of absence of event dependence. It allows broader application of the PERR approach.

*Word count*: 2639

**Plain Language Summary**

Prior event rate ratio (PERR) is an approach to control confounding in observational studies. It assumes that occurrence of outcome events does not influence later event rate, or in other words, absence of "event dependence". We propose using the conditional frailty method for implementation of PERR in the presence of event dependence and evaluate its performance by simulation. In simulation we found that the proposed method successfully reduced the bias usually to smaller than 10%. We demonstrate the use of the method with a study of emergency department visit rate and palliative care in patients with advanced cancer. While conventional time-to-event regression analysis estimated higher hazard of emergency department visits among palliative care recipients (HR=3.61, P<0.001), crude PERR estimate and the proposed PERR estimate were 1.45 (P=0.22) and 1.22 (P=0.57), respectively. The proposed bias reduction method mitigates the impact of violation of the PERR assumption of absence of event dependence. It allows broader application of the PERR approach.

# 1. Introduction

Real-world evidence plays a significant role in the evaluation of the effectiveness and safety of medical treatments [1]. Confounding is a major challenge in real-world evidence research. The prior event rate ratio (PERR) is an approach developed for controlling measured or unmeasured confounders [2,3]. It has been applied to the evaluation of a variety of treatments [4-7].

The central idea of PERR is to partition the person-time of persons who were treated with the intervention concerned during a study duration into two periods: prior to or after initiation of treatment. Each treated person is matched to at least one control person who has not received the treatment by the end of the study duration. The treated person's date of treatment initiation is the index date for dividing the matched control's person-time into prior and post periods. The Cox model for analysis of time-to-first-event is used to compare the hazard of the outcome events in the two groups separately in the prior and post period, giving two hazard ratios (HR), $HR_{prior}$ and $HR_{post}$, respectively. Since two Cox models are estimated separately, a person may have a "first" event in the prior period and another "first" event in the post period. The PERR estimate of treatment effect is the ratio of the HRs:

$$\text{PERR HR} = HR_{post}/HR_{prior}.$$

On the assumption that the magnitude of confounding is constant over time, the method cancels out the confounding embedded in $HR_{prior}$ and $HR_{post}$ and therefore PERR HR is an estimate of treatment effect.

We have recently developed an alternative implementation of the PERR approach, by including data from both periods in an extension of the Cox-model for recurrent events analysis, the Andersen-Gill (AG) model [8]. The model includes a main-effect term for treated vs control group, a main-effect term for post vs prior period, and a group-by-period interaction term, where the interaction term has the interpretation of PERR HR.

One of the PERR assumptions is that earlier events do not influence later event rate, i.e. absence of "event dependence". Otherwise, the PERR HR estimate is biased. Previous research has demonstrated how to evaluate the plausibility of this assumption [9]. However, so far there is no proposal on how to control for this bias should there be event dependence. The conditional frailty (CF) method is a strategy already in the time-to-event analysis literature for controlling for the impact of event dependence [10,11]. Based on our formulation of the PERR method as one time-to-event analysis model with a group-by-period interaction term, we propose to incorporate the CF method into the PERR HR estimation.

We apply this method to the use of palliative care among patients with advanced cancer. Clinical and healthcare researchers have hypothesized that palliative care (PC) can reduce acute healthcare utilization and healthcare costs in patients with advanced illness [12,13], but evidence of such benefits is scant [12,14]. Some researchers suggested that the differences in trials and real-world context tended to lead to under-estimation of PC benefits in trials [15]. Real-world data is therefore expected to play an important role in the evaluation of PC. We will use simulation to demonstrate the value of incorporating CF method into PERR HR estimation. Then we will illustrate the method with an evaluation of incidence rate of emergency department (ED) visits in relation to uptake of PC in patients with advanced cancer.

## 2. Materials and Methods

*2.1. Statistical Models*

Estimation of PERR HR

The formulation based on AG model with a treatment-by-period interaction term is:

$$\lambda_i(t) = \lambda_0(t) \exp(\beta_1 \times trt_i + \beta_2 \times post_i + \beta_3 \times trt_i \times post_i), \quad (1)$$

where $\lambda_0(t)$ is the unspecified baseline hazard, $trt_i=1$ if the i-th participant was ever treated during the observation period and $trt_i=0$ otherwise, $post_i=1$ for person-time in the post period and $post_i=0$ otherwise. Person-times in the prior period stop at the treatment/index times. The start time of the post period is not reset to zero [16,17]. This model can be estimated by standard software such as Stata's *stcox* program. With this formula, $HR_{prior} = \exp(\beta_1)$, $HR_{post} = \exp(\beta_1+\beta_3)$, and PERR HR $= \exp(\beta_3)$. We call this the $PERR_{AG}$ HR. One advantage of this approach to the estimation of PERR HR is that it allows an extension to control the impact of event dependence.

When the PERR assumption of absence of event dependence is violated, we embed the conditional frailty (CF) method into the model:

$$\lambda_{ij}(t) = \omega_i \lambda_{0j}(t) \exp(\beta_1 \times trt_i + \beta_2 \times post_i + \beta_3 \times trt_i \times post_i), \quad (2)$$

where $\lambda_{0j}(t)$ is the unspecified baseline hazard function for the j$^{th}$ event stratum (j=1, 2, 3, …, m). The CF method controls the impact of event dependence by making comparison only within strata of persons who have the same event history. However, the stratification itself generates a selection bias. As such, the person-level frailty (random-effect) term, $\omega_i$, is introduced to mitigate this selection bias [10,11]. The quantity $\exp(\beta_3)$ in equation (2) is referred to as PERR$_{CF}$ HR. This model can be estimated by software such as Stata's *strmcure* program [18].

Assessment of PERR model assumption

We have previously proposed and illustrated the use of the dynamic random-intercept model (DRIM) to evaluate the assumption of absence of event dependence [9]. This requires partitioning follow-up time into discrete intervals and fitting a special form of random-effect model as described by Aitkin and Alfo [19,20], which can be implemented in software such as Stata's *gsem* program [9]. The association between event occurrence in an interval and that in the subsequent interval is quantified as an odds ratio.

## 2.2. Simulation

We evaluated the CF method for controlling event dependence in PERR by simulation. We considered two forms of event dependence, transient and constant, and in two directions, positive and negative. In transient and constant forms of event dependence, the occurrence of an event modified the subsequent log(event rate) by a multiplicative factor of $\psi \exp(-0.5t)$ and $\zeta$, respectively, where $t$ was time since the previous event and $\exp(-0.5t)$ indicates exponential decay of the effect of the previous event. In positive event dependence, $0 < \psi$ and $0 < \zeta$, and vice versa. Simulation parameters such as sample size were generated to resemble the PC study in the next section. 500 replicates were used for each scenario. Relative bias of HR estimate (R. Bias), root-mean-square-error (RMSE) of log(HR) estimates, and coverage probability of 95% confidence interval (CP) are presented. Details of the simulation scenarios/parameters and procedures are available in Online Supplementary Materials 1. Stata codes for the simulation are available from https://github.com/Emily-MXM/PERR-ED.

## 2.3. COMPASS study

The Cost of Medical Care of Patients with Advanced Serious Illness in Singapore (COMPASS) is a cohort study of 600 adult patients with stage IV solid cancer, recruited between 2016 and 2018 from National Cancer Centre Singapore (NCCS) and National University Hospital System [21,22]. The study is approved by SingHealth Centralized Institutional Review Board (2015-2781) and National University of Singapore Institutional Review Board (S-20-155).

Healthcare utilization and demographic data were obtained from electronic health records, last updated (censored) at end of 2021. In addition, the patients were interviewed every 3 months till death or end of 2021. The interviews included the Functional Assessment of Cancer Therapy-General (FACT-G) questionnaire. For illustration purpose, we analysed ED visits in relation to community-based palliative care (PC for brevity), including home and day care and regardless of frequency/duration of utilization. The dataset included dates of initiation of PC and ED visits of 600 patients with stage 4 cancer till December 2021, of whom 2 who had missing values in key covariates and 23 patients who had received PC before enrolment were excluded.

Before we applied the PERR approach, we first evaluated the degree of event dependence by partitioning the follow-up time into 90-day intervals and using DRIM with covariates adjustment for time-constant covariates including age at study entry, gender, cancer type, MediFund status and education, and two time-varying covariates, Physical Well-being and Functional Well-being scores of FACT-G.

The PC recipients were matched to control persons at 1:1 ratio using type of cancer and MediFund status (indicator of financial difficulty) as matching variables. We began with sorting the PC recipients according to their time at initiation of PC and non-PC recipients according to their end time of follow-up ascendingly. For each PC recipient we searched for a non-recipient as a control person subject to the condition that the candidate must be still under follow-up at the time the corresponding treated person initiated PC. If there were more than one candidate for the control, we randomly chose one.

The PERR literature has recommended restricting study duration to mitigate the possibility of changes in magnitude of confounding over time, the absence of which is another key assumption of the PERR method. In COMPASS, the median time between initiation of PC and death was about 70 days. We restricted the PERR analysis of ED visits to up to 70 days before and up to 70 days after treatment/index time.

For comparison and illustration purpose we also included time-to-event analysis (not PERR) using the AG model for recurrent event, with PC as a time-varying exposure variable and age at study entry, gender, cancer type, MediFund status and education as time-constant covariates and Physical Well-being and Functional Well-being scores of FACT-G as time-varying covariates.

## 3. Results

*3.1. Simulation results*

Tables 1 shows the performance of $PERR_{AG}$ and $PERR_{CF}$; the former did not and the latter did control the bias arising of event dependence. In the presence of event dependence, the $PERR_{AG}$ HRs were biased. When the previous events constantly or transiently reduced the event rate, i.e. negative event dependence ($\zeta < 0$ or $\psi < 0$), $PERR_{AG}$ HRs were biased towards the null value, by up to about 26% among the scenarios considered. When previous events constantly or transiently increased later event rate, i.e. positive event dependence ($\zeta > 0$ or $\psi > 0$), $PERR_{AG}$ HRs were biased away from the null value, by up to about 96% among the scenarios. The coverage probability of the 95% CI was as low as 41% in the presence of event dependence.

In contrast, the $PERR_{CF}$ approach reduced relative bias as well as RMSE in the presence of event dependence. The median of the absolute values of R. Bias was 5%. The coverage probability also improved to near the nominal level although it can still be somewhat lower than the target.

In the scenarios with no event dependence but the CF method was applied, there was about +/- 10% bias in the estimates; the average estimates were about 0.55 and 1.8 when the true HRs were 0.5 and 2.0, respectively. While the bias was mild if CF was used when it was not needed, it is preferable to evaluate the assumption of no event dependence before applying the CF method.

Further simulation results varying the baseline hazard pattern are available in Online Supplementary Materials 2. The findings are roughly similar to the above.

*3.2. Emergency department visits and palliative care*

<u>Sample description</u>

Among the 575 patients, 185 did and 390 did not received PC during the study period. The matched PERR analysis cohort consisted of a total of 370 patients, including 185 PC recipients and 185 controls.

Table 2 shows the number of ED visits, person-time and ED visit rates in the prior and post periods, with up to 70 days per period. In the post period, the rate ratio was 4.14/0.66 = 6.27. It seemed to indicate higher hazard in the PC group. However, the rate ratio was 2.79/0.65 = 4.29 in the prior period to begin with, showing that it was those who had higher ED visit rates who tended to receive PC later on. There was evidence of strong confounding.

<u>Evaluation of PERR model assumption</u>

In the full cohort (N=575), DRIM showed that occurrence of ED visits in the previous time interval was associated with higher event rate in the subsequent time interval, with odds ratio 1.35 (95% CI 1.01 to 1.81; P=0.042). Event dependence was present.

<u>Time-to-event analysis</u>

For comparison and illustration purpose, time-to-event regression analysis (AG model, N=575) with PC as a time-varying exposure variable gave HR 3.61 (95% CI 2.83 to 4.61) despite adjustment for the group of time-constant and time-varying covariates aforementioned (Table 3).

PERR analysis

Table 3 also shows the estimates from PERR method. Without controlling event dependence, $PERR_{AG}$ HR was 1.45 (95% CI 0.80 to 2.62). When adjusted for event dependence using the CF method, the $PERR_{CF}$ HR was 1.22 (95% CI 0.61 to 2.43). The estimates were in line with the observation from simulation that positive event dependence would lead to bias away from the null value.

## 4. Discussion

The PERR method is a powerful approach for real-world evidence research if its model assumptions are met, including the assumption of no event dependence. Method for empirical assessment of the assumption is available [9]. However, there has been no guidance on what to do when the assumption is violated.

The CF method has been used in conventional time-to-event analysis to mitigate the bias arising from event dependence. For example, in vaccine studies it is used to address the impact of negative event dependence arising from naturally acquired partial immunity associated with infectious disease episodes [11]. In this article we proposed to incorporate the CF method into the estimation of PERR HR and evaluated this strategy by simulation. In the case study of patients with advanced cancer, conventional time-to-event regression analysis showed statistically significant elevation of ED visit rate among PC recipients, with HR=3.61, which was opposite to clinical expectation. This finding was obtained despite adjustment for time-constant confounders from electronic health records and time-varying confounders from 3-monthly survey assessment. The reasons of ED visits are complex and may relate to family/social situations and informal caregiver's characteristics. Electronic health records are typically patient-centric and do not capture such important confounders. Survey assessments of confounders may not be comprehensive and assessments at discrete time-intervals necessitate measurement errors especially in the advanced illness context where patient/caregiver situations may change quickly within a time-interval. As such, analytic methods like multivariable regression analysis or propensity score that are based on assumption of no unmeasured covariates is unlikely to control for all sources of bias. PERR is a promising alternative that is minimal in data requirement. The PERR estimation without adjustment for event dependence (PERR$_{AG}$) gave a HR of 1.45. Despite statistical non-significance, the point estimate was likely exaggerated (away from null value) due to the presence of positive event dependence as indicated by dynamic random-intercept modelling. Applying the CF method to adjust for event dependence (PERR$_{CF}$) gave a HR of 1.22, in line with expectation of the direction of adjustment for positive event dependence.

The PERR requires other assumptions, including non-differential mortality [23,24]. In time-to-event analysis, joint modelling of mortality and recurrent events can be used to control the bias arising from non-differential morality [25,26]. Another advantage of using the AG model as shown in equation (1) to estimate PERR HR is that it makes the joint modelling with mortality readily usable. However, to simultaneously handle the biases arising from violation of both assumptions of no event dependence and non-differential mortality involves another level of complexity. Although conceptually it is possible to incorporate joint modelling of mortality into the CF method, i.e. equation (2), it is a new problem in terms of estimation algorithm and software development. It is a limitation of the present work that we have not provided a more general method to simultaneously handle both problems. Furthermore, the study of ED visit rates in relation to PC is not conclusive as we have not considered whether there was bias arising from differential mortality. Further development in the methodological research and application to PC studies are needed.

## 5. Conclusion

The conditional frailty method can be incorporated into prior event rate ratio estimation when the assumption of event dependence is violated. Observed elevation of ED visit rates was in part due to confounding, which can be mitigated by PERR method, and in part due to event dependence, which can be mitigated by the proposed application of CF method.

**Ethics Approval**

The study is approved by SingHealth Centralized Institutional Review Board (2015-2781) and National University of Singapore Institutional Review Board (S-20-155).

**Data sharing and data availability statement**

Stata codes for the simulation reported are available from https://github.com/Emily-MXM/PERR-ED. COMPASS study data are not publicly available due to restrictions on distribution of electronic health record data. Data are however available from the authors upon reasonable request and permission of SingHealth and National University Health Systems and institutional review board approval from the requesting institution.

**Funding**

The methodological work was supported by the National Medical Research Council, Singapore (MOH-001487). The COMPASS study was funded by Singapore Millennium Foundation (2015-SMF-0003) and Lien Centre for Palliative Care (LCPC-IN14-0003).

**Declaration of Interests**

None.

**Disclaimer**

Any opinions, findings and conclusions or recommendations expressed in this material are those of the authors and do not reflect the views of Ministry of Health / National Medical Research Council, Singapore.

**ORCID**

Xiangmei Ma https://orcid.org/0000-0001-6526-1226
Chetna Malhotra https://orcid.org/0000-0002-5380-0525
Eric Finkelstein https://orcid.org/0000-0001-6443-9686
Yin Bun Cheung https://orcid.org/0000-0003-0517-7625

**References**

1. Corrigan-Curay J, Sacks L, Woodcock J. Real-world evidence and real-world data for evaluating drug safety and effectiveness. *Journal of American Medical Association* 2018; 320(9):867-868. doi: 10.1001/jama.2018.10136.

2. Tannen RL, Weiner MG, Xie D. Use of primary care electronic medical record database in drug efficacy research on cardiovascular outcomes: comparison of database and randomised controlled trial findings. *British Medical Journal* 2009; 338: b81. doi: 10.1136/bmj.b81.

3. Rodgers LR, Dennis JM, Shields BM, et al. Prior event rate ratio adjustment produced estimates consistent with randomized trial: a diabetes case study. *Journal of Clinical Epidemiology* 2020; 122: 78-86. doi: 10.1016/j.jclinepi.2020.03.007.

4. Hamilton F, Arnold D, Henley W, Payne RA. Aspirin reduces cardiovascular events in patients with pneumonia: a prior event rate ratio analysis in a large primary care database. *European Respiratory Journal* 2021; 57(2): 2002795. doi: 10.1183/13993003.02795-2020.

5. Scott J, Jones T, Redaniel MT, et al. Estimating the risk of acute kidney injury associated with use of diuretics and renin angiotensin aldosterone system inhibitors: A population based cohort study using the clinical practice research datalink. *BMC Nephrology* 2019; 20(1): 481. doi: 10.1186/s12882-019-1633-2.

6. Streeter AJ, Rodgers LR, Hamilton F, et al. Influenza vaccination reduced myocardial infarctions in United Kingdom older adults: a prior event rate ratio study. *Journal of Clinical Epidemiology* 2022; 151: 122-131. doi: 10.1016/j.jclinepi.2022.06.018.

7. Streeter AJ, Rodgers LR, Masoli J, et al. Real-world effectiveness of pneumococcal vaccination in older adults: Cohort study using the UK Clinical Practice Research Datalink. *PLoS One* 2022; 17(10): e0275642. doi: 10.1371/journal.pone.0275642.

8. Ma X, Yang GM, Zhuang QY, Cheung YB. Strategy to control biases in prior event rate ratio method, with application to palliative care in patients with advanced cancer. *arXiv* 2024 [pre-print]; arXiv:2412.17879. doi: 10.48550/arXiv.2412.17879.

9. Cheung YB, Ma X, Mackenzie G. Two assumptions of the prior event rate ratio approach for controlling confounding can be evaluated by self-controlled case series and dynamic random intercept modeling. *Journal of Clinical Epidemiology* 2024; 175:111511. doi: 10.1016/j.jclinepi.2024.111511.

10. Box-Steffensmeier JM, De Boef S. Repeated events survival models: the conditional frailty model. *Statistics in Medicine* 2006; 25(20):3518-33. doi: 10.1002/sim.2434.

11. Cheung YB, Ma X, Lam KF, Milligan P. Estimation of the primary, secondary and composite effects of malaria vaccines using data on multiple clinical malaria episodes. *Vaccine* 2020; 38(32):4964-4969. doi: 10.1016/j.vaccine.2020.05.086.

12. DiMartino LD, Weiner BJ, Mayer DK, et al. Do palliative care interventions reduce emergency department visits among patients with cancer at the end of life? A systematic review. *Journal of Palliative Medicine* 2014; 17(12):1384-1399. doi: 10.1089/jpm.2014.0092.

13. Hsu HS, Wu TH, Lin CY, et al. Enhanced home palliative care could reduce emergency department visits due to non-organic dyspnea among cancer patients: a retrospective cohort study. *BMC Palliative Care* 2021; 20(1):42. doi: 10.1186/s12904-021-00713-6.

14. Kavalieratos D, Corbelli J, Zhang D, et al. Association between palliative care and patient and caregiver outcomes: A Systematic Review and Meta-analysis. *Journal of American Medical Association* 2016; 316(20):2104-2014. doi: 10.1001/jama.2016.16840.

15. Gaertner J, Siemens W, Meerpohl JJ, et al. Effect of specialist palliative care services on quality of life in adults with advanced incurable illness in hospital, hospice, or community settings: systematic review and meta-analysis. *British Medical Journal* 2017; 357: j2925. doi: 10.1136/bmj.j2925.

16. Lin NX, Henley WE. Prior event rate ratio adjustment for hidden confounding in observational studies of treatment effectiveness: a pairwise Cox likelihood approach. *Statistics in Medicine* 2016; 35(28): 5149-5169. doi: 10.1002/sim.7051.

17. Cleves M. Analysis of multiple failure-time data with Stata. *Stata Technical Bulletin* 2000; 9(49): 338-349.

18. Xu Y, Cheung YB. Frailty models and frailty mixture models for recurrent event times. *Stata Journal* 2015; 15: 135-154 . doi: 10.1177/1536867X1501500109.

19. Aitkin M, Alfo M. Regression models for binary longitudinal responses. *Statistics and Computing* 1998; 8: 289–307. doi: 10.1023/A:1008999824193.

20. Skrondal A, Rabe-Hesketh S. Handling initial conditions and endogenous covariates in dynamic/transition models for binary data with unobserved heterogeneity. *Journal of the Royal Statistical Society, Series C* 2014; 63(2): 211-237. doi: 10.1111/rssc.12021.

21. Teo I, Singh R, Malhotra C, et al. Cost of Medical Care of Patients with Advanced serious Illness in Singapore (COMPASS): prospective cohort study protocol. *BMC Cancer* 2018; 18(1):459. doi: 10.1186/s12885-018-4356-z.

22. Malhotra C, Bundoc F, Chaudhry I, et al. A prospective cohort study assessing aggressive interventions at the end-of-life among patients with solid metastatic cancer. *BMC Palliat Care* 2022; 21(1):73. doi: 10.1186/s12904-022-00970-z.

23. Uddin MJ, Groenwold RH, van Staa TP, et al. Performance of prior event rate ratio adjustment method in pharmacoepidemiology: a simulation study. *Pharmacoepidemiology and Drug Safety* 2015; 24(5): 468-77. doi: 10.1002/pds.3724.

24. Thommes EW, Mahmud SM, Young-Xu Y, et al. Assessing the prior event rate ratio method via probabilistic bias analysis on a Bayesian network. *Statistics in Medicine* 2020; 39(5):639-659. doi: 10.1002/sim.8435.

25. Liu L, Wolfe RA, Huang X. Shared frailty models for recurrent events and a terminal event. *Biometrics* 2004; 60(3):747-56. doi: 10.1111/j.0006-341X.2004.00225.x.

26. Rondeau V, Mathoulin-Pelissier S, Jacqmin-Gadda H, et al. Joint frailty models for recurring events and death using maximum penalized likelihood estimation: application on cancer events *Biostatistics* 2007; 8(4):708-21. doi: 10.1093/biostatistics/kxl043.

Table 1. Simulation results on the performance of crude PERR method (PERR$_{AG}$) and the PERR adjusted for event dependence using the CF method (PERR$_{CF}$); N = 600 in pre-matched cohort. [1]

| Event dependence pattern [2,3] | Parameters | Mean N | Mean no. of events | PERR$_{AG}$ | | | PERR$_{CF}$ | | |
|---|---|---|---|---|---|---|---|---|---|
| | | | | R. Bias (%) | RMSE | CP (%) | R. Bias (%) | RMSE | CP (%) |
| Negative, constant | HR = 0.5; $\zeta$ = −1.0 | 326 | 691 | 19.5 | 0.224 | 80.6 | 6.1 | 0.188 | 92.8 |
| Negative, constant | HR = 2.0; $\zeta$ = −1.0 | 326 | 920 | -18.9 | 0.254 | 63.2 | -3.2 | 0.170 | 91.8 |
| Negative, transient | HR = 0.5; $\psi$ = −1.0 | 326 | 573 | 25.6 | 0.269 | 69.2 | 12.6 | 0.217 | 88.8 |
| Negative, transient | HR = 2.0; $\psi$ = −1.0 | 326 | 719 | -26.2 | 0.342 | 41.0 | -10.9 | 0.216 | 86.6 |
| Positive, constant | HR = 0.5; $\zeta$ = 1.0 | 326 | 414 | -19.7 | 0.567 | 86.0 | -3.5 | 0.325 | 91.0 |
| Positive, constant | HR = 2.0; $\zeta$ = 1.0 | 326 | 1026 | 95.5 | 0.715 | 64.2 | -5.7 | 0.250 | 93.6 |
| Positive, transient | HR = 0.5; $\psi$ = 1.0 | 326 | 487 | -8.0 | 0.503 | 91.4 | 3.3 | 0.307 | 89.6 |
| Positive, transient | HR = 2.0; $\psi$ = 1.0 | 328 | 773 | 23.6 | 0.431 | 93.4 | -2.0 | 0.257 | 91.8 |
| None | HR = 0.5 | 326 | 542 | 2.8 | 0.213 | 94.4 | 9.2 | 0.218 | 92.0 |
| None | HR = 2.0 | 326 | 872 | 2.5 | 0.206 | 93.4 | -10.5 | 0.249 | 78.6 |

[1] R. Bias = relative bias; RMSE = root-mean-square-error; CP = coverage probability of 95% confidence interval.
[2] Negative and positive event dependence: events reduce and increase later event rate, respectively.
[3] Constant: each event permanently changes later log(even rate) by $\zeta$; transient: each event has an impact on later log(event rate) that decays over time by $\psi \exp(-0.5t)$.

Table 2. Incidence of emergency department visits (visits/person-years) by palliative care (PC) groups and periods, Singapore, 2016 to 2021 (N=370).

| Period | PC | Non-PC |
|--------|----|----|
| Post | 4.14 (108/26.1) | 0.66 (23/34.9) |
| Prior | 2.79 (94/33.7) | 0.65 (22/33.7) |

Table 3. Prior event rate ratio analysis and time-to-event analysis of emergency department visit rates in relation to palliative care uptake in patients with advanced cancer, Singapore 2016 to 2021 (N=370).

| Analysis method | HR | 95% CI | P |
|---|---|---|---|
| Time-to-event | 3.61 | 2.83 to 4.61 | <0.001 |
| $PERR_{AG}$ | 1.45 | 0.80 to 2.62 | 0.217 |
| $PERR_{CF}$ | 1.22 | 0.61 to 2.43 | 0.568 |

**Online Supplementary Materials.** Supplemental procedures on simulation and supplemental simulation results.

## *S1. Supplemental procedures on simulation of scenarios with event dependency*

Generation of time-to-event data with event dependency

Let $\tau_i$ be the censoring time of the i-th participant, $i = 1,2,\ldots,n$.

Time to treatment of the i-th participant, $t_{trt,i}$, is generated from the hazard function of a Weibull distribution:

$$h_{trt}(t) = 0.5t^{-0.5}u_i\exp(c_0 + 0.5X_i + 0.5Z_i), \quad (1)$$

where $X_i$ is an unobserved binary confounder with $Prob(X_i = 1) = 0.5$, $Z_i$ is an observed binary covariate with $Prob(Z_i = 1) = 0.5$, and $u_i$ is an independent random Gamma variates with mean 1 and variance 0.1. Define the time-varying treatment indicator to be

$$P_i(t) = 1 \text{ if } t_{trt,i} < t \leq \tau_i; \text{otherwise } P_i(t) = 0.$$

The participant is eligible for inclusion as a treated person if $t_{trt,i} < \tau_i$; otherwise the participant is eligible as a control person.

Time to outcome events is generated from the hazard function of a Weibull distribution:

$$\lambda_i(t) = kt^{k-1}w_i\exp(\beta_0 + \beta P_i(t) + E_i(t) + 0.5X_i + 0.5Z_i), \quad \text{for } k > 0 \quad (2)$$

where $E_i(t)$ is the effect of event dependence, and $w_i$ is an independent random Gamma variates with mean 1 and variance $\sigma_\omega^2$. We consider two patterns of $E_i(t)$, representing constant and decaying effects of event history, respectively:

$$E_i(t) = \zeta ln[N_i(t\text{-}) + 1], \quad (3)$$

$$E_i(t) = \sum_{j=1}^{N_i(t\text{-})} \psi e^{-0.5t} I(t_{ij} < t), \quad (4)$$

where $N_i(t\text{-})$ is the number of events prior to time $t$ and $t_{ij}$ is the time of the j-th event of the i-th person.

Data generation begins with generating observed follow-up time, $\tau_i \sim \text{Uniform}(a,b)$. Then $\{t_{ij}: j = 1,2,\ldots,n_i\}$ are iteratively generated by thinning method:

i. Set $t_{i0} = 0$; set the initial value of $t^*$ be 0.001 for person $i$ and $j = 1$.

ii. Draw a random number $R \sim \text{Exp}(\bar{\lambda}_{ij})$, where $\bar{\lambda}_{ij}$ is a fixed value for person $i$ while having $j-1$ events such that $\lambda_i(t) \leq \bar{\lambda}_{ij}$, for $\forall t > t_{i,j-1}$. Here, $\bar{\lambda}_{ij} = k t_{sup,i}^{k-1} w_i \exp(\beta_0 + \beta 1_{\beta>0} + E_{sup,j} + 0.5 X_i + 0.5 Z_i)$, and $t_{sup,i} = \tau_i$ if $k > 1$, otherwise, $t_{sup,i} = 0.001$; if equation (3) is applied to equation (2) and $\zeta > 0$ then $E_{sup,j} = \zeta \ln(j)$, and if equation (4) is applied to equation (2) and $\psi > 0$ then $E_{sup,j} = (j-1)\psi$, otherwise, $E_{sup,j} = 0$.

iii. Update $t^* = t^* + R$. If $t^* \geq \tau_i$, then let $t_{ij} = \tau_i$ be time at censoring. The iteration for person $i$ stops here.

iv. If $t^* < \tau_i$, generate a random number $V \sim \text{Unifrom}(0,1)$.

v. If $V > \frac{\lambda_i(t^*)}{\bar{\lambda}_{ij}}$, repeat the process from step (ii). If $V \leq \frac{\lambda_i(t^*)}{\bar{\lambda}_{ij}}$, let $t^* = t_{ij}$ be an event time, and update $j = j + 1$; repeat the process from step (ii) to generate $t_{i,j+1}$. $\lambda_i(t^*)$ has time-varying covariates $P_i(t^*)$ and $E_i(t^*)$ as defined in equation (2).

Main set of simulation scenarios/parameters

Pre-match sample size is set at 600 persons, similar to the case study. For implementation of PERR, match treated persons to control persons at 1:1 ratio on $Z_i$. The sample size for the matched PERR dataset is subject to chance and scenario parameters and usually between 250 and 420. Each scenario is evaluated with 500 replicates.

For correction of bias due to event dependency, we use PERR method with the conditional frailty model ($\text{PERR}_{\text{CF}}$), we report $\text{PERR}_{\text{AG}}$ and $\text{PERR}_{\text{CF}}$ with $k = 0.8, 1.0$ or $1.2$ and positive or negative event dependency. Since the simulated dataset may become sparse at higher-event strata, we pool the $j$-th and later events to form one stratum in the CF model, where $j$ is the value determined at the 95th percentile of the number of events across all persons. Unless otherwise specified, we set the parameters to be $c_0 = -2.0$, $\sigma_\omega^2 = 0.5$, and $\tau_i \sim \text{Uniform}(1, 3)$, we will examine two different treatment effects, HR = 0.5 or 2. The value of $\beta_0$ is set to be varied from -1.7 to 0.5 in different scenarios.

Statistics results presented in Tables

We show the corresponding results in tables in terms of R. Bias (%), RMSE and CP(%), where R. Bias (%) = $\frac{\widehat{HR} - \text{True HR}}{\text{True HR}} \times 100\%$ is the relative bias of mean of $HR_i$ estimates, $\widehat{HR} = \frac{1}{500}\sum_{i=1}^{500} HR_i$, in percentage obtained from the 500 replicates, and RMSE = $\sqrt{\frac{1}{500}\sum_{i=1}^{500}[\log(HR_i) - \log(\text{True HR})]^2}$ is the mean of root-mean-square-error of $\log(HR_i)$ estimates, and CP(%) is coverage probability of 95% confidence interval.

## S2. Supplemental simulation results

Table S1. Simulation results on the performance of crude PERR method (PERR$_{AG}$) and the PERR adjusted for event dependence using the CF method (PERR$_{CF}$); N = 600 in pre-matched cohort and $k = 1.2$ in equation (2). [1]

| Event dependence pattern [2,3] | Parameters | Mean N | Mean no. of events | PERR$_{AG}$ | | | PERR$_{CF}$ | | |
|---|---|---|---|---|---|---|---|---|---|
| | | | | R. Bias (%) | RMSE | CP (%) | R. Bias (%) | RMSE | CP (%) |
| Negative, constant | HR = 0.5; $\zeta = -1.0$ | 328 | 719 | 22.9 | 0.245 | 73.4 | 7.9 | 0.203 | 90.8 |
| Negative, constant | HR = 2.0; $\zeta = -1.0$ | 326 | 994 | -19.8 | 0.267 | 62.8 | -1.9 | 0.185 | 90.4 |
| Negative, transient | HR = 0.5; $\psi = -1.0$ | 328 | 618 | 29.5 | 0.290 | 61.8 | 16.6 | 0.233 | 83.8 |
| Negative, transient | HR = 2.0; $\psi = -1.0$ | 326 | 795 | -28.2 | 0.370 | 32.0 | -12.5 | 0.241 | 80.8 |
| Positive, constant | HR = 0.5; $\zeta = 1.0$ | 326 | 578 | -24.6 | 0.774 | 78.6 | -7.1 | 0.394 | 83.8 |
| Positive, constant | HR = 2.0; $\zeta = 1.0$ | 326 | 2191 | 196.4 | 1.084 | 20.2 | 3.3 | 0.312 | 86.0 |
| Positive, transient | HR = 0.5; $\psi = 1.0$ | 326 | 500 | -7.1 | 0.580 | 89.0 | 13.9 | 0.320 | 89.0 |
| Positive, transient | HR = 2.0; $\psi = 1.0$ | 328 | 908 | 28.7 | 0.490 | 91.2 | -9.1 | 0.271 | 90.6 |
| None | HR = 0.5 | 324 | 578 | 2.6 | 0.238 | 93.8 | 12.2 | 0.238 | 90.0 |
| None | HR = 2.0 | 326 | 1004 | 3.6 | 0.238 | 92.2 | -11.7 | 0.283 | 76.2 |

[1] R. Bias = relative bias; RMSE = root-mean-square-error; CP = coverage probability of 95% confidence interval.
[2] Negative and positive event dependence: events reduce and increase later event rate, respectively.
[3] Constant: each event permanently changes later log(even rate) by $\zeta$; transient: each event has an impact on later log(event rate) that decays over time by $\psi \exp(-0.5t)$.

Table S2. Simulation results on the performance of crude PERR method (PERR$_{AG}$) and the PERR adjusted for event dependence using the CF method (PERR$_{CF}$); N = 600 in pre-matched cohort and $k = 1.0$ in equation (2). [1]

| Event dependence pattern [2,3] | Parameters | Mean N | Mean no. of events | PERR$_{AG}$ | | | PERR$_{CF}$ | | |
|---|---|---|---|---|---|---|---|---|---|
| | | | | R. Bias (%) | RMSE | CP (%) | R. Bias (%) | RMSE | CP (%) |
| Negative, constant | HR = 0.5; $\zeta = -1.0$ | 326 | 702 | 21.8 | 0.241 | 75.4 | 8.6 | 0.212 | 89.0 |
| Negative, constant | HR = 2.0; $\zeta = -1.0$ | 328 | 966 | -20.0 | 0.269 | 58.6 | -2.9 | 0.172 | 93.8 |
| Negative, transient | HR = 0.5; $\psi = -1.0$ | 326 | 591 | 30.7 | 0.298 | 62.0 | 18.1 | 0.234 | 85.4 |
| Negative, transient | HR = 2.0; $\psi = -1.0$ | 326 | 757 | -27.8 | 0.363 | 31.6 | -12.6 | 0.234 | 82.6 |
| Positive, constant | HR = 0.5; $\zeta = 1.0$ | 328 | 501 | -24.8 | 0.692 | 77.6 | -7.8 | 0.370 | 87.8 |
| Positive, constant | HR = 2.0; $\zeta = 1.0$ | 326 | 1834 | 188.4 | 1.062 | 44.4 | 10.7 | 0.324 | 86.2 |
| Positive, transient | HR = 0.5; $\psi = 1.0$ | 328 | 531 | -7.3 | 0.553 | 91.4 | 9.4 | 0.326 | 86.8 |
| Positive, transient | HR = 2.0; $\psi = 1.0$ | 326 | 940 | 29.3 | 0.495 | 91.4 | -9.6 | 0.270 | 88.4 |
| None | HR = 0.5 | 328 | 563 | 4.3 | 0.208 | 95.8 | 12.9 | 0.225 | 91.4 |
| None | HR = 2.0 | 328 | 942 | 4.9 | 0.211 | 93.0 | -9.3 | 0.250 | 80.8 |

[1] R. Bias = relative bias; RMSE = root-mean-square-error; CP = coverage probability of 95% confidence interval.
[2] Negative and positive event dependence: events reduce and increase later event rate, respectively.
[3] Constant: each event permanently changes later log(even rate) by $\zeta$; transient: each event has an impact on later log(event rate) that decays over time by $\psi \exp(-0.5t)$.